\documentclass[12pt]{iopart}

\usepackage{amsmath}
\usepackage{iopams}  
\usepackage{graphicx}
\usepackage{harvard}
\usepackage{hyperref}
\newcommand{\degree}{\ensuremath{^\circ}}

\begin{document}
\title{Light scattering from an isotropic layer between uniaxial crystals}

\author{E S Thomson$^1$, L A Wilen$^{1,2}$ and J S Wettlaufer$^{1,3,4}$}

\address{$^1$ Department of Geology and Geophysics, Yale University, New Haven, CT, 06520, USA}
\address{$^2$ Unilever Research and Development, Trumball, CT, 06611, USA}
\address{$^3$ Department of Physics, Yale University, New Haven, CT, 06520, USA}
\address{$^4$ Nordic Institute for Theoretical Physics, Roslagstullsbacken 23, SE-106 91 Stockholm, Sweden}
\ead{erik.thomson@yale.edu}
\begin{abstract}
We develop a model for the reflection and transmission of plane waves by an isotropic layer sandwiched between two uniaxial crystals of arbitrary orientation.   In the laboratory frame, reflection and transmission coefficients corresponding to the principal polarization directions in each crystal are given explicitly in terms of the $\hat{c}$-axis and propagation directions.  The solution is found by first deriving explicit expressions for reflection and transmission amplitude coefficients for waves propagating from an arbitrarily oriented uniaxial anisotropic material into an isotropic material.   By  combining these results with \possessivecite{Lekner1991} earlier treatment of waves propagating from isotropic media to anisotropic media and employing a matrix method we determine a solution to the general form of the multiple reflection case.  The example system of a wetted interface between two ice crystals is used to contextualize the results.

\end{abstract}



\section{Introduction}

Since the time of Descartes scientists have taken an interest in the physical properties of ice \cite{Dash2006}.  For example, \citeasnoun{Tyndall1856} observed harvested ice in some detail, but it was not for almost another hundred years, when \citeasnoun{Nakaya1954} began cataloging observations of snowflakes, that well-controlled laboratory growth experiments on single crystals began.  More recently, motivated by the fact that commonly occurring environmental temperatures span the triple point of ice, a strong understanding of the thermodynamics and phase behavior of polycrystalline ice near its melting temperature has been developed.  It is known that near the melting temperature, an interconnected network of liquid water exists within the polycrystalline solid \cite{Nye1973}.  Under hydrostatic conditions, veins of water separate the boundaries between three crystals and join in nodes where four grains meet.  The Gibbs-Thomson and impurity effects are responsible for the presence of this liquid \cite{Dash2006}, which is observed using optical microscopy techniques \cite{Mader1992,Walford1987}.  Additional water structures, such as water lenses, are observed in the presence of non-hydrostatic stresses \cite{Nye1991}.  Less well understood is what happens away from these junctions in the planar interface between two single crystals.  While disorder is expected at the molecular scale, it is predicted that a dopant, such as salt, can induce the formation of a thick ($>10$ nm), essentially bulk, water film \cite{Benatov2004}.  This phenomenon of \emph{interfacial melting} could have important implications for ice's electrical and mechanical properties and impurity redistribution in glaciers and polar ice \citeaffixed{Rempel2002}{e.g.}.  More generally, it may occur in other polycrystalline materials. Recent predictions of interfacial melting at ice grain boundaries have motivated an experimental search \cite{Thomson2005a} to detect the water layer using an optical reflection technique.  This has led us to consider the theoretical formalism for wave reflection and transmission in an anisotropic/isotropic/anisotropic layered system; specifically when the anisotropic media are uniaxial crystals. 

Uniaxial crystals are scientifically well studied materials owing to their ubiquity in nature and their many technical applications, including use as elements in optical systems.  Theoretical treatments of light propagation in these crystals have focused on reflection from surfaces, and the internal propagation through layered structures.  Previous studies have used $4 \times 4$ or $2 \times 2$ matrix methods to solve the general problem of light propagation through birefringent networks, where solutions are given in principal axes coordinate frames \cite{Yeh1979,Yeh1982}.  These treatments leave the reader to solve the eigenvalue problem associated with transforming to a laboratory coordinate frame of experimental relevance. In other studies only special $\hat{c}$-axis orientations are considered \citeaffixed{Yeh1982}{e.g.} or multiple internal reflections are ignored \citeaffixed{Gu1993}{e.g.}.  Still other methods find solutions at a single interface and are, due to their form, difficult to extrapolate to multiple interfaces or multiple reflections \cite{Stamnes1977,Zhang1996}. These existing theoretical studies are not ideally suited to experimental applications; to be of utility solutions must be valid for arbitrary crystallographic orientations, incidence angle, and propagation direction, all measured in the laboratory frame.

Our study approaches the problem from the perspective of the specific experimental setting described above; a wetted interface between two uniaxial crystals of arbitrary orientation.  We begin by revisiting the problem of plane wave propagation in a uniaxial crystal using modal decomposition, the approach by which \citeasnoun{Lekner1991} determined the reflection and transmission amplitudes for a plane wave entering an anisotropic medium from an isotropic medium.  The relevance of \possessivecite{Lekner1991} important earlier work to our study requires that we begin by reviewing his results in some detail.  Here we generalize those results in order to analyze the reverse situation;  where the wave passes from an anisotropic to an isotropic region.  We explicitly determine the reflection and refraction amplitude coefficients in terms of the orientation of the $\hat{c}$-axis with respect to the laboratory axes and the optical constants of the materials.  Consequently, we are able to solve for all of the relevant amplitude coefficients associated with an isotropic layer sandwiched between uniaxial crystals.  This enables us to construct a matrix method to model the Fabry-Perot effect of multiple reflections from the isotropic layer.  Extensions of the theory may also be applicable to more general birefringent systems, but here it is of particular interest at the interface between two grains in water ice.  To illustrate this we present clear examples of how the generalized theory can be used in comparison with light reflection experiments. 

\section{Anisotropic Optical Theory}
\begin{figure}
\centering
\includegraphics[scale=.6]{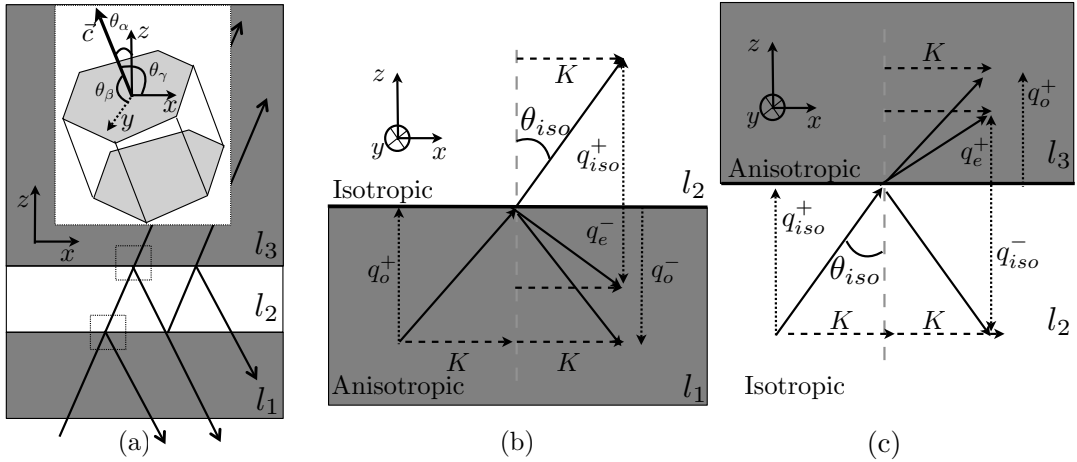}
\caption{(a) The general geometry of the three layer system.  Plane wave propagation is illustrated with rays, enlarged schematics of the boxed interface regions are shown in (b) and (c) and in $l_3$ an example crystal is inset.  The $\hat{c}$-axis direction of the example crystal is labeled with respect to the laboratory frame of reference.  Here $\theta_\alpha$ is the angle between $\hat{c}$ and $\hat{x}$, $\theta_\beta$ is the angle between $\hat{c}$ and $\hat{y}$,  and $\theta_\gamma$ is the angle between $\hat{c}$ and $\hat{z}$.  (b) Schematic of an \emph{o} wave incident on an anisotropic/isotropic boundary, where $K$ is the preserved tangential component of all wave vectors, $q$'s are the normal ($z$) components of the wave vectors, and $\theta_{iso}$ is the angle of the transmitted wave vector.  An analogous schematic of an incident \emph{e} wave could be drawn.  (c) An \emph{s} or \emph{p} wave, they are co-linear, incident on the isotropic/anisotropic interface.}  
\label{fig:anisis}
\end{figure}

The problem of interest, illustrated schematically in \fref{fig:anisis}(a), is that of plane wave propagation in a three layer system, an isotropic layer ($l_2$), bounded by uniaxial crystals ($l_1$ and $l_3$).  Within the isotropic layer a wave's polarization can be decomposed in the usual way to be parallel ($p$) and perpendicular ($s$) to the plane of incidence.  In the anisotropic material the principal components are parallel and perpendicular to the optical axis of the material; these are referred to as the extraordinary ($e$) and ordinary ($o$) modes.  To characterize the system completely we describe the plane wave propagation within each layer, in addition to the reflection and refraction at the boundaries between the media.  

In the laboratory frame of reference the reflecting surfaces are \emph{xy} planes and $z$ is the normal; the \emph{zx} plane is chosen as the plane of incidence.  The electric field is denoted $\mathbf{E}=[E_x,E_y,E_z]e^{i(qz+Kx-\omega t)}$.  No $y$ dependence exists due to translational symmetry in the $y$ direction.  Continuity of the tangential component of $\mathbf{E}$ demands that $K$ is common in all media while the normal component of the wave vector ($q$) will depend on the state of polarization, propagation direction and specific medium of propagation.  For example the value of $q$ corresponding to an ordinary ray propagating in the -$z$ direction is denoted by $q_o^-$.  The angular frequency is $\omega$, thereby defining a wavevector $k=\omega /c$.  Referring to \fref{fig:anisis}(b) for waves incident from an anisotropic media onto an anisotropic/isotropic boundary reflected \emph{o} and \emph{e} waves and colinear, transmitted \emph{s} and {p} waves result.  Conversely (\Fref{fig:anisis}(c)), for waves incident from an isotropic media onto an isotropic/anisotropic boundary \emph{o} and \emph{e} waves are transmitted and colinear \emph{s} and \emph{p} waves are reflected.  For the three layer system, a given $K$ determines unique $q_{iso}$ and $\theta_{iso}$ values within the sandwiched isotropic layer.  Alternatively, if one was interested in a single angle of incidence in the uniaxial crystal the problem could be solved iteratively using different $K$ values for the \emph{o} and \emph{e} polarizations.

\subsection{Plane Wave Propogation within a Uniaxial Crystal}
\citeasnoun{Lekner1991} employed a normal mode analysis to study wave propagation in uniaxial crystals.  Using an orthogonal coordinate transformation, expressed in terms of direction cosines, he first expressed the dielectric tensor of a uniaxial crystal in the laboratory frame of reference.  He then explicitly determined the ordinary ($\mathbf{E}^o$) and extraordinary ($\mathbf{E}^e$) electric field vectors, in addition to the wave vector's \emph{z}-components for propagation within the crystal, in terms of $K$, $k$, and the c-axis orientation, specified by the unit vector $\hat{c}=[\alpha,\beta,\gamma]$; where $\alpha= \cos{\theta_\alpha}$, $\beta= \cos{\theta_\beta}$, $\gamma= \cos{\theta_\gamma}$ (see \fref{fig:anisis}(a)). 
\begin{eqnarray}
\label{eq:Eo} \mathbf{E}^{o^\pm}= N_o^\pm[-\beta q_o^\pm,\alpha q_o^\pm - \gamma K,\beta K],\\
\label{eq:Ee} \mathbf{E}^{e^\pm}=N_e^\pm[\alpha q_o^2-\gamma q_e^\pm K, \beta \varepsilon_ok^2,\gamma (\varepsilon_ok^2-q_e^{\pm^2})-\alpha q_e^\pm K].
\end{eqnarray}
Here $q_o$ and $q_e$ are the normal components of the ordinary and extraordinary wave vectors, $\varepsilon _o=n_o^2$ is the ordinary dielectric constant, and $N_o$ and $N_e$ are normalization constants. The expressions for the \emph{z}-component of the wave vectors are
\begin{eqnarray}
\label{eq:qo} q_o^\pm=\pm \sqrt{\varepsilon_ok^2-K^2}\;\;\;\text{and}\\
\label{eq:qe} q_e^\pm=\frac{\pm \sqrt{D}-\alpha \gamma K \Delta \varepsilon}{\varepsilon_o+\gamma^2\Delta \varepsilon},
\end{eqnarray}
where in all cases the signs ($\pm$) correspond to the direction of beam propagation with respect to the \emph{z}-axis.  The quantity $D$, in $q_e^\pm$, is given by 
\begin{equation}
\label{eq:d} D=\varepsilon_o[\varepsilon_e(\varepsilon_o+\gamma^2\Delta \varepsilon)k^2-(\varepsilon_e-\beta^2\Delta \varepsilon)K^2],
\end{equation}
where $\varepsilon_e$ is the extraodinary dielectric constant and $\Delta\varepsilon = \varepsilon_e-\varepsilon_o$. 

\possessivecite{Lekner1991} analysis is general to plane wave propagation within uniaxial materials and provides a foundation for investigating reflection and refraction at interfaces with such materials.   

\subsection{Lekner's Amplitude Coefficients}

\citeasnoun{Lekner1991} goes on to calculate reflection and transmission coefficients for \emph{s} and \emph{p} waves incident from an isotropic media onto an isotropic/anistropic interface.  Because the tangential components are preserved across the boundary he focuses on the $z$-dependence of the electric field.  The \emph{z}-dependencies of the incident, reflected, and transmitted electric fields for the \emph{s} polarization are,
\begin{eqnarray} 
\label{eq:incs} \text{incident:} && {E}^{in}=e^{iq_{iso}^+z}[0,1,0], \nonumber \\
\label{eq:refs}\text{reflected:} && {E}^{ref}=r_{sp}e^{i q_{iso}^{-}z}[\cos{\theta_{iso}},0,-\sin{\theta_{iso}}]+r_{ss}e^{iq_{iso}^{-}z}[0,1,0],\\
\label{eq:transs} \text{transmitted:}\;  && {E}^{tr}=t_{so}e^{iq_{o}^+z}[E_x^{o^+},E_y^{o^+},E_z^{o^+}]+t_{se}e^{iq_{e}^+z}[E_x^{e^+},E_y^{e^+},E_z^{e^+}]. \nonumber
\end{eqnarray}
At the interface the electromagnetic waves are subject to the following boundary conditions implied by Maxwell's equations; continuity of $E_x$, $E_y$, $\partial E_x/\partial z - iKE_z$, and $\partial E_y/ \partial z$, where the subscripts $x,y,z$ refer to the vector components.  Applying these boundary conditions at the reflecting plane ($z=0$) leads to four equations that \citeasnoun{Lekner1991} solved for the four unknown intensity coefficients: $r_{ss}$,$r_{sp}$,$t_{so}$, and $t_{se}$. In Appendix A \eref{eq:rss}-\eref{eq:tse} we summarize those results with small changes correcting apparent typographical errors in the original publication.  The incident, reflected, and transmitted \emph{p} waves are
\begin{eqnarray} 
\label{eq:incs} \text{incident:} && {E}^{in}=e^{iq_{iso}^+z}[\cos{\theta_{iso}},0,-\sin{\theta_{iso}}], \nonumber \\
\label{eq:refs}\text{reflected:} && {E}^{ref}=r_{pp}e^{i q_{iso}^{-}z}[\cos{\theta_{iso}},0,-\sin{\theta_{iso}}]+r_{ps}e^{iq_{iso}^{-}z}[0,1,0],\\
\label{eq:transs} \text{transmitted:}\;  && {E}^{tr}=t_{po}e^{iq_{o}^+z}[E_x^{o^+},E_y^{o^+},E_z^{o^+}]+t_{pe}e^{iq_{e}^+z}[E_x^{e^+},E_y^{e^+},E_z^{e^+}]. \nonumber
\end{eqnarray}
Again, application of the four boundary conditions at the interface leads to four amplitude coefficients, $r_{pp}$,$r_{ps}$,$t_{po}$, and $t_{pe}$, also presented in Appendix A \eref{eq:rpp}-\eref{eq:tpe}.

When considering these results for transmission from isotropic media to anisotropic media it is critical to realize that within the anisotropic material the ray direction (i.e., the Poynting vector) for the extraordinary mode differs from the wave vector direction.  This subtlety must be recognized to verify simple test cases of reflection and refraction.  For more discussion regarding ray direction see \citeasnoun{Lekner1991}.  

\subsection{Anisotropic to Isotropic Interface}

We analyze the reverse incidence, when \emph{o} and \emph{e} waves are incident from an anisotropic media onto an anisotropic/isotropic interface, in a similar manner.  For the ordinary wave the \emph{z}-dependencies of the electric fields are
\begin{eqnarray} 
\label{eq:inco} \text{incident:} && {E}^{in}=e^{iq_o^+z}[E_x^{o^+},E_y^{o^+},E_z^{o^+}], \nonumber \\
\label{eq:refo}\text{reflected:} && {E}^{ref}=r_{oo}e^{i q_{o}^{-}z}[E_x^{o^-},E_y^{o^-},E_z^{o^-}]+r_{oe}e^{iq_{e}^{-}z}[E_x^{e^-},E_y^{e^-},E_z^{e^-}],\\
\label{eq:transo} \text{transmitted:}\;  && {E}^{tr}=t_{os}e^{iq_{iso}^+z}[0,1,0]+t_{op}e^{iq_{iso}^+z}[\cos{\theta _{iso}},0,-\sin{\theta_{iso}}]. \nonumber
\end{eqnarray}
Applying the boundary conditions to the ordinary wave at the reflecting plane ($z=0$) the above general expressions for the incident, reflected, and transmitted waves \eref{eq:refo} yield a system of four equations;
\begin{eqnarray}
\label{eq:Ex} E_x^{o^+}+r_{oo}E_x^{o^-}+r_{oe}E_x^{e^-}-t_{op}\cos{\theta_{iso}}=0,\\
\label{eq:Ey} E_y^{o^+}+r_{oo}E_y^{o^-}+r_{oe}E_y^{e^-}-t_{os}=0,\\
\label{eq:dEx,Ez} q_o^{+}E_x^{o^+}+q_{o}^{-}r_{oo}E_x^{o^-}+q_{e}^{-}r_{oe}E_x^{e^-}-q_{iso}^+t_{op}\cos{\theta_{iso}}-\nonumber\\
\;\;\;K(E_z^{o^+}+r_{oo}E_z^{o^-}+r_{oe}E_z^{e^-}+t_{op}\sin{\theta_{iso}})=0, \\
\label{eq:dEy} q_o^{+}E_y^{o^+}+q_{o}^{-}r_{oo}E_y^{o^-}+q_{e}^{-}r_{oe}E_y^{e^-}-q_{iso}^+t_{os}=0,
\end{eqnarray}
and four unknown amplitude coefficients $r_{oo}$,$r_{oe}$, $t_{os}$ and $t_{op}$ for the ordinary wave.
Solving this homogeneous system of equations for the unknown amplitude coefficients provides expressions for  $r_{oo}$,$r_{oe}$, $t_{os}$ and $t_{op}$ shown in their complete form in Appendix B \eref{eq:roo}-\eref{eq:top}.  

The extraordinary wave's intensity coefficients can be found in a manner analogous to those for the ordinary wave. Again we begin with expressions for the \emph{z}-dependence of the extraordinary electric field:
\begin{eqnarray}
\label{eq:ince} \text{incident:} && {E}^{in}=e^{iq_e^+z}[E_x^{e^+},E_y^{e^+},E_z^{e^+}], \nonumber \\
\label{eq:refe} \text{reflected:} && {E}^{ref}=r_{ee}e^{iq_{e}^-z}[E_x^{e^-},E_y^{e^-},E_z^{e^-}]+r_{eo}e^{iq_{o}^-z}[E_x^{o^-},E_y^{o^-},E_z^{o^-}],\\
\label{eq:transe} \text{transmitted:}\; && {E}^{tr}= t_{es}e^{iq_{iso}^+z}[0,1,0]+t_{ep}e^{iq_{iso}^+z}[\cos{\theta _{iso}},0,-\sin{\theta_{iso}}] .\nonumber
\end{eqnarray}
We find solutions for the intensity coefficients of an incident extraordinary beam ($r_{ee}$,$r_{eo}$, $t_{es}$ and $t_{ep}$) as we did previously; see \eref{eq:ree}-\eref{eq:tep}.  Thus, the magnitudes of the derived amplitude coefficients \eref{eq:roo}-\eref{eq:tep} are fully determined by completing the normal mode analysis substitutions \eref{eq:Eo}-\eref{eq:d}.

A limiting case provides some verification of the now explicit amplitude coefficients for an incident \emph{o} wave \eref{eq:roo}-\eref{eq:top} presented in Appendix B.  If the $\hat{c}$-axis of the crystal is in the plane of incidence and the incident \emph{o} wave is entirely perpendicularly polarized with respect to the plane of incidence (i.e. $[E_x^o,E_y^o,E_z^o]=[0,1,0]$) the reflection and transmission amplitude coefficients reduce to the Fresnel equations for perpendicular polarization \citeaffixed{Born1965}{e.g.}:
\begin{eqnarray}
\label{eq:Fres,r} r_{oo}\rightarrow r_s \rightarrow \frac{q_o^+-q_{iso}^+}{q_{iso}^+-q_{o}^-}\rightarrow \frac{q_o^+-q_{iso}^+}{q_{iso}^+ +q_{o}^+}\;\;\;\text{and}\\
\label{eq:Fres,t} t_{os}\rightarrow t_s \rightarrow \frac{q_o^{+}-q_{o}^-}{q_{iso}^+-q_{o}^-}\rightarrow \frac{2q_o^+}{q_{iso}^++q_{o}^+},
\end{eqnarray}
remembering $q_o^+=-q_o^-$.  As expected the other coefficients ($r_{oe}$, $t_{op}$) vanish.  In contrast, the extraordinary amplitude coefficients can be modeled using isotropic theory for the \emph{p} polarization as long as it is recognized that this introduces an effective index of refraction that is a function of incident angle \cite{Born1965},
\begin{equation}
\label{eq:neff}n_{eff}=\frac{n_on_e}{\sqrt{n_o^2\sin{\theta_i^\prime}+n_e^2\cos{\theta_i^\prime}}}.
\end{equation}
%
\begin{figure}
\centering
\includegraphics[scale=.6]{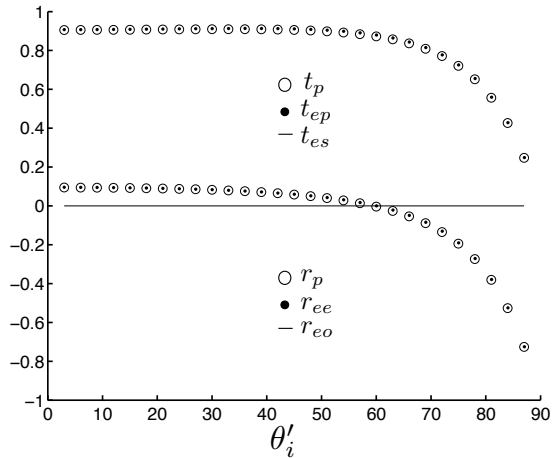}
\caption{Extraordinary reflection and transmission coefficients for a beam incident on the basal plane ($[\theta_\alpha,\theta_\beta,\theta_\gamma]=[90\degree,90\degree,180\degree]$) compared with isotropic theory.  Here the Fresnel equations for $p$ polarization solutions are used with $n_{eff}$ \eref{eq:neff}.  Circles are isotropic theory and the points (inside circles) are the full theory.   The line at zero represents the cross term coefficients, which for this crystal orientation are always zero.  The physical constants used are $n_o=1.1$, $n_e=1.2$ and $n_{iso}=1.33$.}
\label{fig:coeffs}
\end{figure}
\noindent Here $\theta_i^\prime$ is the wavevector incident angle on the boundary and $n_o$ and $n_e$ are the ordinary and extraordinary indices of refraction.  It also must be noted that within the crystal the wave vector and electric field are not necessarily perpendicular.  Therefore, the angle used to compute the Fresnel coefficients must be that of the Poynting vector (the ray direction), while Snell's law must be solved using the wave vector direction.  Figure \ref{fig:coeffs} illustrates the agreement between the coefficients for an extraordinary beam incident on a basal plane and the isotropic Fresnel equations.

Now that we have investigated plane wave propagation within each region and across each boundary, individually; we return to the three layer system.  By collecting the expressions for each of the relevant reflection and transmission amplitude coefficients it is possible to construct a matrix formulation for the propagation of light through the uniaxial network.

\section{Matrix Method}
\label{sec:Matform}

Similar to the Jones matrix formalism, $2\times 2$ matrices can be used to describe reflection and refraction at interfaces with uniaxial materials \citeaffixed{Yeh1982,Abdulhalim1999}{e.g.}.  Rather than rotation matrices, as in the Jones formulation, here the matrix elements are the relevant amplitude coefficients.  The diagonal elements represent reflection or refraction of like polarization, while the off-diagonal elements represent the mixing of polarization states.  Each interface is represented by two independent matrices, one representing reflection, the other refraction.   For example, reflection at an anisotropic/isotropic interface can be written in matrix form as 
\begin{equation}\label{eqn:E1r}
\mathbf{E}_{1r}=\begin{pmatrix} r_{oo} & r_{eo}\\
r_{oe} & r_{ee} \end{pmatrix} \binom{E_{i}^o}{E_{i}^e}\equiv R_1\mathbf{E}_i ,
\end{equation}
where both the reflected wave, $\mathbf{E}_{1r}\equiv(E_{1r}^o , E_{1r}^e)$, and the incident wave, $\mathbf{E}_i\equiv(E_{i}^o, E_{i}^e)$, will have ordinary and extraordinary components.  The reflection matrix $R_1$ is composed of the amplitude coefficients representing the interface which incorporate the properties of the anisotropic material.  The phase shift acquired by waves that travel some distance through a uniaxial material can be accounted for using diagonal propagation matrices, such as  
\begin{equation}\label{eq:P1}
P_1= \begin{pmatrix} e^{-i\delta_o} & 0\\
0 & e^{-i\delta_e}  \end{pmatrix}.
\end{equation}
The phase factors for the \emph{o} and \emph{e} waves are given by $\label{eq:deltaO} \delta_o=\Lambda _o\sqrt{q^2_o+K^2}$ and $\delta_e=\Lambda_e\sqrt{q^2_e+K^2}$, where $\Lambda_o$ and $\Lambda_e$ are their respective path lengths within the crystal.

This matrix approach for treating different interfaces substantially simplifies the analyses of light scattering associated with layered materials.  However, it is important to note, when modeling multiple layers and/or reflections careful attention must be paid to maintaining a consistent coordinate system.  In the next section we use this approach to examine the specific example of an isotropic layer sandwiched between anisotropic layers.

\section{Anisotropic/Isotropic/Anisotropic Layering - Application to Grain Boundaries}
\label{sec:GBA}
\begin{figure}
\centering
\includegraphics[scale=.9]{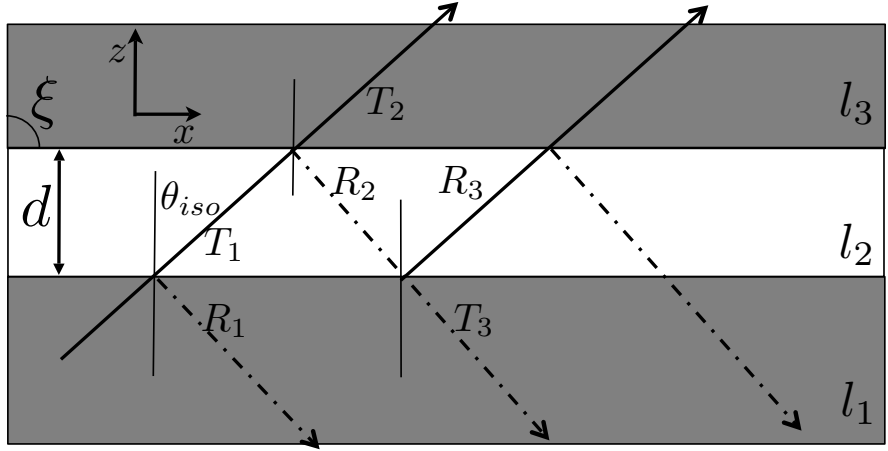}
\caption{Schematic of multiple reflections off of a grain boundary interface.  Different interfaces are labeled with the relevant amplitude coefficient matrices for reflection and transmission. The isotropic film ($l_2$) between the crystals, $l_1$ and $l_3$, has thickness $d$. } 
\label{fig:Fulltrace}
\end{figure}

As discussed in the introduction, a primary motivation of this study is to better understand the phase behavior of ice and other polycrystalline materials near their melting temperatures.  One proposed experimental method for characterizing the grain boundary is to measure a reflected laser beam's intensity as a function of the thermodynamic variables: temperature, crystal orientation, and impurity concentration \cite{Thomson2005a}.  However, data gathered from such an experiment can only be interpreted accurately with a theoretical model that includes the anisotropy of the system.  The combination of our results with those of \citeasnoun{Lekner1991} leads to precisely that type of model; of an isotropic layer sandwiched between uniaxial crystals (\Fref{fig:Fulltrace}).  While here we focus on ice and water to make a connection to our experiment, the theoretical model applies to any such geometry and anisotropy.

First we ignore any optical path length within the ice crystals, and simply consider the problem as if the incident beam were generated and the reflected beam observed at the interface of $l_1$ and $l_2$ (\Fref{fig:Fulltrace}).  Propagation away from the interface can be accounted for using \eref{eq:P1}.  Following the matrix approach of \Sref{sec:Matform} the transmission and reflection coefficients for each interface can be formulated;  
\begin{eqnarray}
\label{eqn:T1}T_1\equiv \begin{pmatrix} t_{op} & t_{ep} \\ t_{os} & t_{es} \end{pmatrix},T_2\equiv \begin{pmatrix} t_{po}' & t_{so}' \\ t_{pe}' & t_{se}' \end{pmatrix},T_3\equiv \begin{pmatrix} t_{po} & t_{so} \\ t_{pe} & t_{se} \end{pmatrix},\\
\label{eqn:R2} R_2\equiv \begin{pmatrix} r_{pp}' & r_{sp}' \\ r_{ps}' & r_{ss}' \end{pmatrix},  R_3\equiv \begin{pmatrix} r_{pp} & r_{sp} \\ r_{ps} & r_{ss} \end{pmatrix},
\end{eqnarray} 
and $R_1$ remains as previously defined by \eref{eqn:E1r}.  Primed amplitude coefficients are used to denote those associated with the $l_2/l_3$ boundary.  For now we ignore the transmitted signal, which does not apply to our experiment.  For this model we define the page to be in the \emph{xz} plane with positive up (\emph{z}) and to the right (\emph{x}), and are careful to insure that the polarization vector direction is consistent in each layer. Therefore, in the unprimed amplitude coefficient solutions of $T_3$ and $R_3$, obtained by solving \eref{eq:rss}-\eref{eq:tpe}, $\theta_{iso}$ is substituted with $-\theta_{iso}$, and the negative normal mode solutions from \eref{eq:Eo}-\eref{eq:qe} are used.  

The waves of experimental interest are the initial and each subsequent reflection ($\mathbf{E}_{1r},\mathbf{E}_{2r},\mathbf{E}_{3r}$...etc.), and can be expressed as a function of the original incoming field as was done in \eref{eqn:E1r} for the $\mathbf{E}_{1r}$ term.   Writing down the first few reflections the pattern is evident;
\begin{eqnarray}
\label{eqn:E2r} \mathbf{E}_{2r}=T_3R_2T_1\mathbf{E}_ie^{i(wt-\delta)},\\
\label{eqn:E3r} \mathbf{E}_{3r}=T_3R_2R_3R_2T_1\mathbf{E}_ie^{i(wt-2\delta)},\\
\label{eqn:E4r} \mathbf{E}_{4r}=T_3R_2R_3R_2R_3R_2T_1\mathbf{E}_ie^{i(wt-3\delta)}.
\end{eqnarray}
These terms include the additional phase contributions arising from the optical path length of each reflection internal to the grain boundary, $\delta=2k n_{iso} d \cos{\theta_{iso}}$ (\Fref{fig:Fulltrace}).  Because the multiple reflections occur within the isotropic medium, the path length depends upon only one angle.  Anisotropy within the intervening layer would further complicate the situation by introducing a second angle of transmission resulting in multiple possible paths within the layer.  In the limit of the superposition of a large number of such reflections, the total reflected field ($\mathbf{E}_{r}^{tot}$) becomes  
\begin{eqnarray}
\label{eqn:Ertot1} \mathbf{E}_{r}^{tot}&=&\mathbf{E}_{1r}+\mathbf{E}_{2r}+\mathbf{E}_{3r}+.....+\mathbf{E}_{nr}\\
\label{eqn:Ertot2} &=&(R_1+\sum_{n=1}^{\infty}T_3R_2\bar{R}^{n-1}e^{-in\delta}T_1)\mathbf{E}_ie^{i\omega t}
\end{eqnarray}
where, $\bar{R}\equiv R_3R_2$.   This expression contains a geometric series of matrices \citeaffixed{Strang1993}{e.g.} and as such can be rewritten, using the identity matrix $I$, as
\begin{equation}\label{eq:Ertot3}
\mathbf{E}_{r}^{tot}=[R_1+T_3R_2e^{-i\delta}(I-\bar{R}e^{-i\delta})^{-1}T_1]\mathbf{E}_ie^{i\omega t}.
\end{equation}
This substitution is valid as long as the absolute values of the eigenvalues of $\bar{R}$ are less than one.  The limiting case of total reflection with no anisotropy, $\bar{R}= \bigl( \begin{smallmatrix} 1&0 \\ 0&1 \end{smallmatrix} \bigr)$, illustrates that any other situation will lead to smaller eigenvalues; less than one.  The bracketed expression in \eref{eq:Ertot3} will remain valid for reflection in any geometry that includes an isotropic sandwich.  We call this  the $n$-reflection matrix ($M_n^{ref}$), 
\begin{multline}\label{eq:Mn}
M_n^{ref}= \left( \begin{array}{c}
{[r_{oo}e^{i2\delta}+(\zeta_4t_{po}+\zeta_5t_{so} -\eta_1r_{oo}) e^{i\delta}+\eta_2(\zeta_3t_{op}+\eta_3r_{oo} +\eta_4t_{os})] }
\\ { [r_{eo}e^{i2\delta}+(\zeta_1t_{po}+\zeta_2t_{so} -\eta_1r_{eo}) e^{i\delta}+\eta_2(\zeta_3t_{ep}+\eta_3r_{eo} +\eta_4t_{es})] }
 \end {array} \right . \\
\left. \begin{array} {c} 
{[r_{oe}e^{i2\delta}+(\zeta_4t_{pe}+\zeta_5t_{se} -\eta_1r_{oe}) e^{i\delta}+\eta_2(\zeta_6t_{op}+\eta_3r_{oe} +\zeta_7t_{os})] }\\
{[r_{ee}e^{i2\delta}+(\zeta_1t_{pe}+\zeta_2t_{se} -\eta_1r_{ee}) e^{i\delta}+\eta_2(\zeta_6t_{ep}+\eta_3r_{ee} +\zeta_7t_{es})]}\end{array} \right)\frac{1}{(e^{i2\delta}-\eta_1e^{i\delta}+\eta_2\eta_3)} .
\end{multline}
In \eref{eq:Mn} the variable quantities within the matrix elements are given in Appendix C.  In an experimental system with a beam propagating through a crystal, the propagation matrices \eref{eq:P1} associated with the optical path length will modify the final result.  

\begin{figure}
\centering 
\includegraphics[scale=.6]{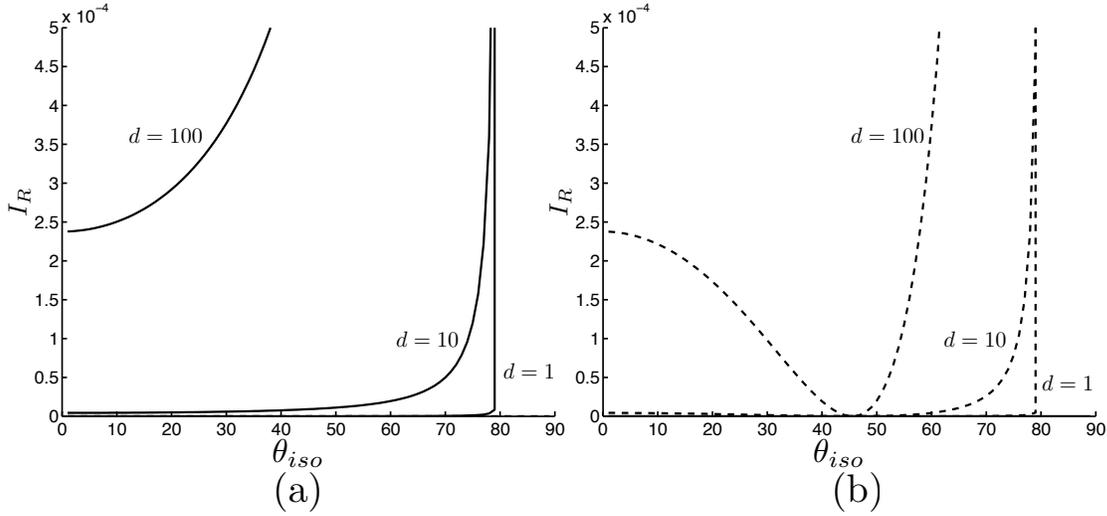}
\caption{Theory for reflected intensity ratio ($I_R$) from the basal planes, $[\theta_\alpha,\theta_\beta,\theta_\gamma]=[90\degree,90\degree,0\degree]$ of each crystal, in an ice-water-ice sandwich.  Intensity ratio is plotted for varying water layer thicknesses, $d=1,10,100$ nm.  In both cases solid curves correspond with reflected \emph{o} polarization and dashed curves with the \emph{e} polarization.  Curves asymptote at the approximate critical angle for total reflection within the water layer ($\approx80\degree$).  (a) The incident beam is \emph{o} polarized $[E_{i}^o,E^{e}_i]=[1,0]$.  Note that in this geometry, for purely ordinary incidence, the extraordinary reflected component is always zero.  (b) The incident beam is \emph{e} polarized $[E^{o}_i,E^{e}_i]=[0,1]$.  Again there is no polarization mixing.}
\label{fig:90900}
\end{figure}

\begin{figure}
\centering
\includegraphics[scale=.6]{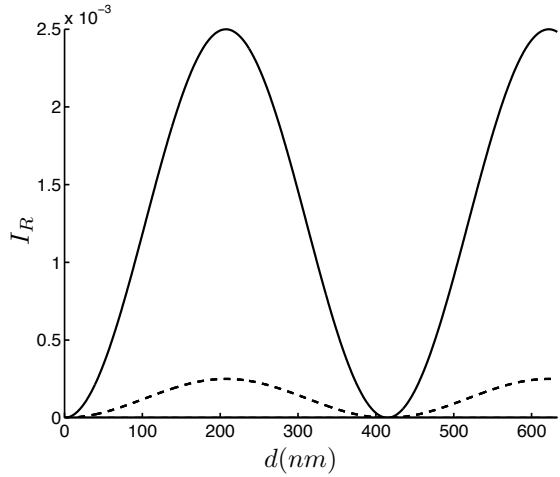}
\caption{Theory for reflected intensity ratio ($I_R$) from the basal planes, $[\theta_\alpha,\theta_\beta,\theta_\gamma]=[90,90,0]$ of each crystal, in an ice-water-ice sandwich as a function of water layer thickness $d$, at an incidence angle of $55\degree$.  The solid curve corresponds to the reflection of an incident wave of purely \emph{o} polarization $[E_{i}^o,E^{e}_i]=[1,0]$ and the dashed curve to the reflection of an  incident wave of \emph{e} polarization $[E^{o}_i,E^{e}_i]=[0,1]$.}
\label{fig:dbasal}
\end{figure}

\begin{figure}
\centering 
\includegraphics[scale=.6]{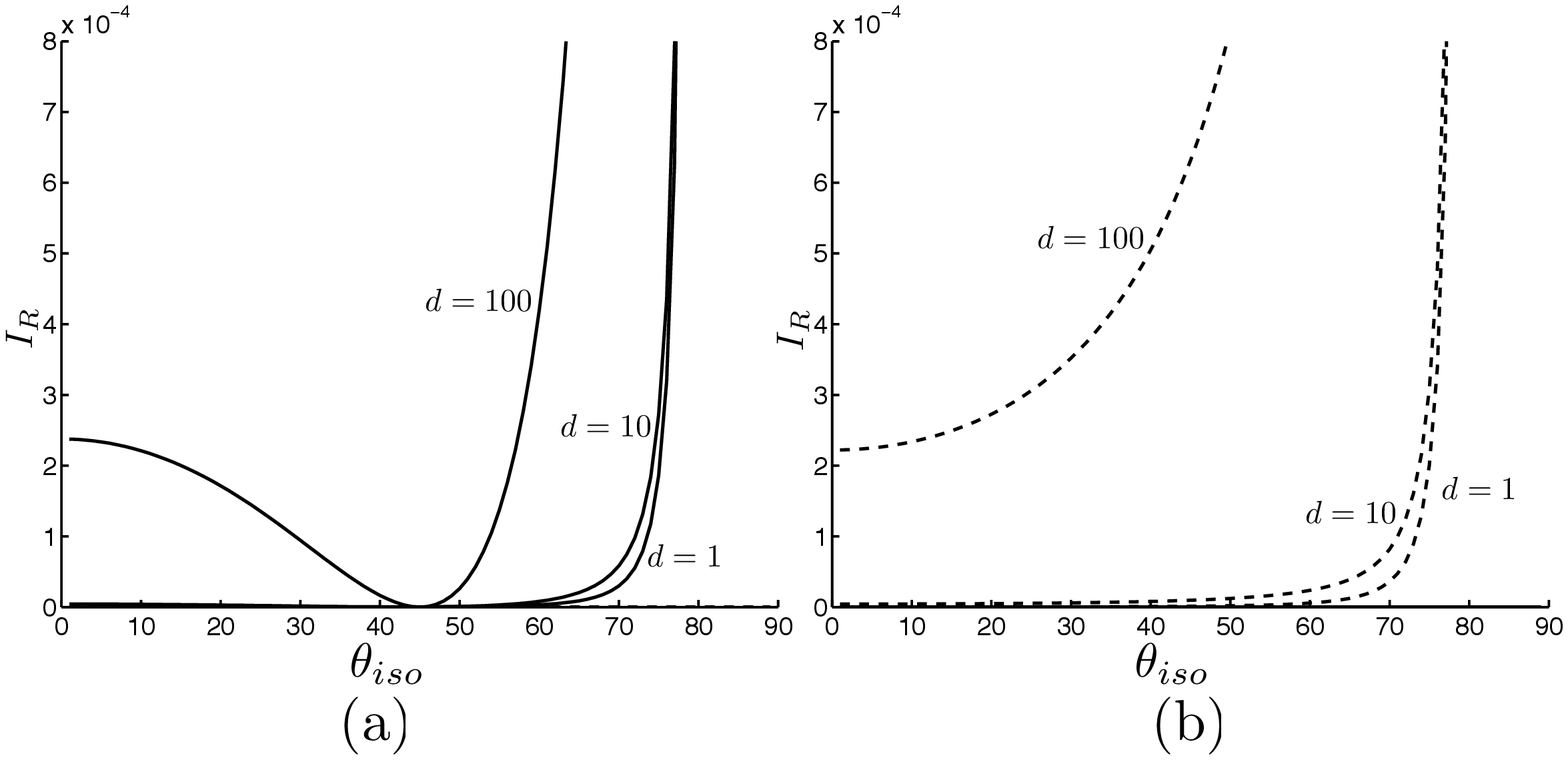}
\caption{Theory for reflected intensity ratio ($I_R$) from an ice-water-ice sandwich. $[\theta_{\alpha}^1,\theta_{\beta}^1,\theta_{\gamma}^1]=[90\degree,0\degree,90\degree]$ and $[\theta_{\alpha}^2,\theta_{\beta}^2,\theta_{\gamma}^2]=[90\degree,90\degree,0\degree]$.  Intensity ratio is plotted for varying water layer thicknesses, $d=1,10,100$ nm.  In both cases solid curves correspond with reflected \emph{o} polarization and dashed curves with the \emph{e} polarization.  (a) The incident beam is \emph{o} polarized $[E^{o}_i,E^{e}_i]=[1,0]$.  (b) The incident beam is \emph{e} polarized $[E^{o}_i,E^{e}_i]=[0,1]$. No polarization mixing occurs in either case.}
\label{fig:90090}
\end{figure}

\begin{figure}
\centering 
\includegraphics[scale=.6]{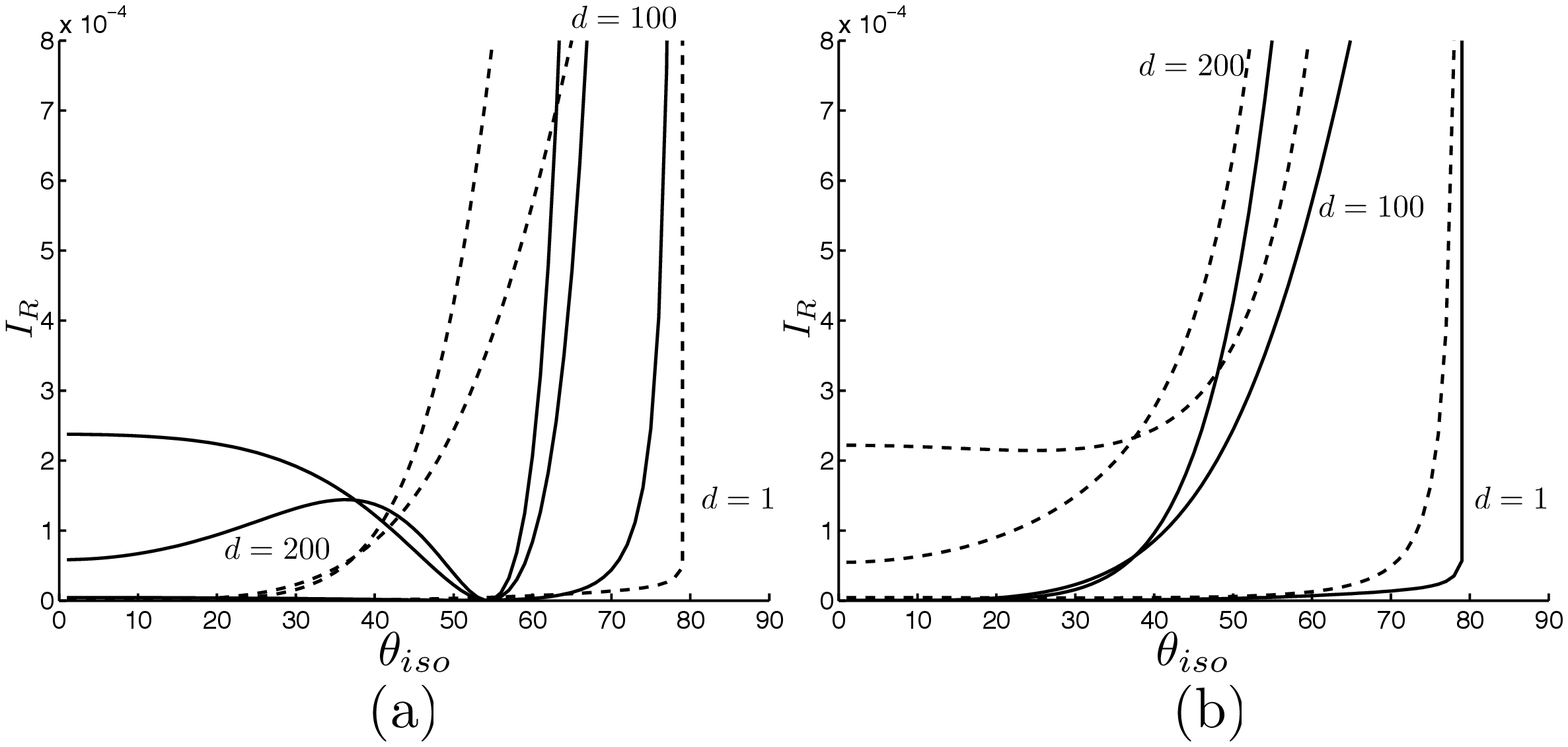}
\caption{Theory for reflected intensity ratio ($I_R$) from an ice-water-ice sandwich. $[\theta_{\alpha}^1,\theta_{\beta}^1,\theta_{\gamma}^1]=[45\degree,45\degree,90\degree]$ and $[\theta_{\alpha}^2,\theta_{\beta}^2,\theta_{\gamma}^2]=[90\degree,90\degree,0\degree]$.  Intensity ratio is plotted for varying water layer thicknesses, $d=1,100,200$ nm.  In both cases solid curves correspond with reflected \emph{o} polarization and dashed curves with the \emph{e} polarization.  (a) The incident beam is \emph{o} polarized $[E^{o}_i,E^{e}_i]=[1,0]$.  (b) The incident beam is \emph{e} polarized $[E^{o}_i,E^{e}_i]=[0,1]$. Polarization mixing occurs in both cases.}
\label{fig:454590}
\end{figure}

\begin{figure}
\centering 
\includegraphics[scale=.6]{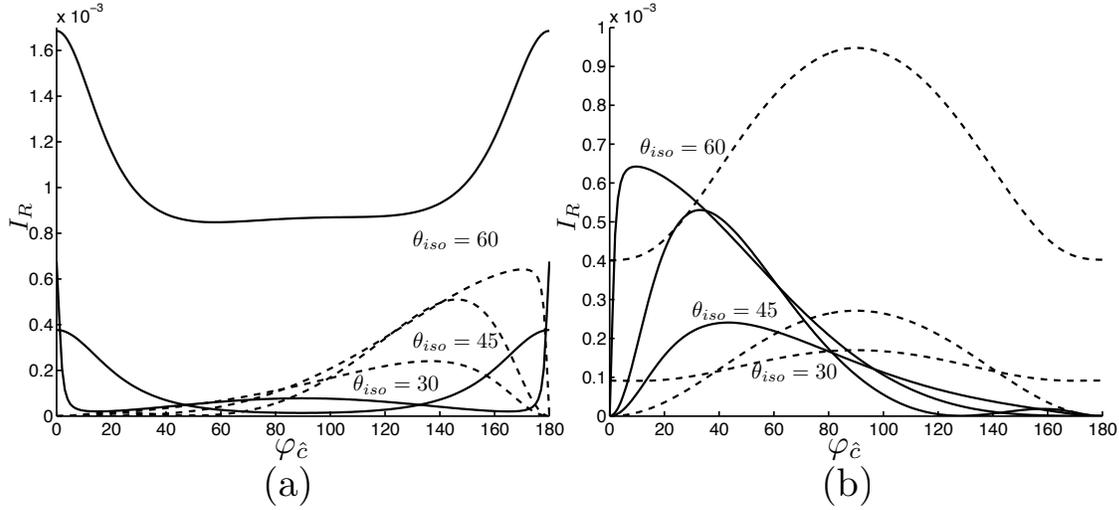}
\caption{Theory for reflected intensity ratio ($I_R$) from an ice-water-ice sandwich for multiple values of $\theta_{iso}$. The $\hat{c}$-axis orientation of $l_1$ is rotated about the $z$-axis.  The angle, $\varphi_{\hat{c}}$, is the projection of $\hat{c}$ onto the $xy$ plane with $\theta_{\gamma}^1=45\degree$ held constant.  Water layer thickness is also a constant, $d=100$ nm, and $[\theta_{\alpha}^2,\theta_{\beta}^2,\theta_{\gamma}^2]=[90\degree,90\degree,0\degree]$.  Again solid curves correspond with reflected \emph{o} polarization and dashed curves with the \emph{e} polarization.  (a) The incident beam is \emph{o} polarized $[E^{o}_i,E^{e}_i]=[1,0]$.  (b) The incident beam is \emph{e} polarized $[E^{o}_i,E^{e}_i]=[0,1]$.  In both cases the curves for $\varphi_{\hat{c}}=180\degree - 360\degree$ are symmetric with what is shown. }
\label{fig:rotcax}
\end{figure}

\begin{figure}
\centering 
\includegraphics[scale=.6]{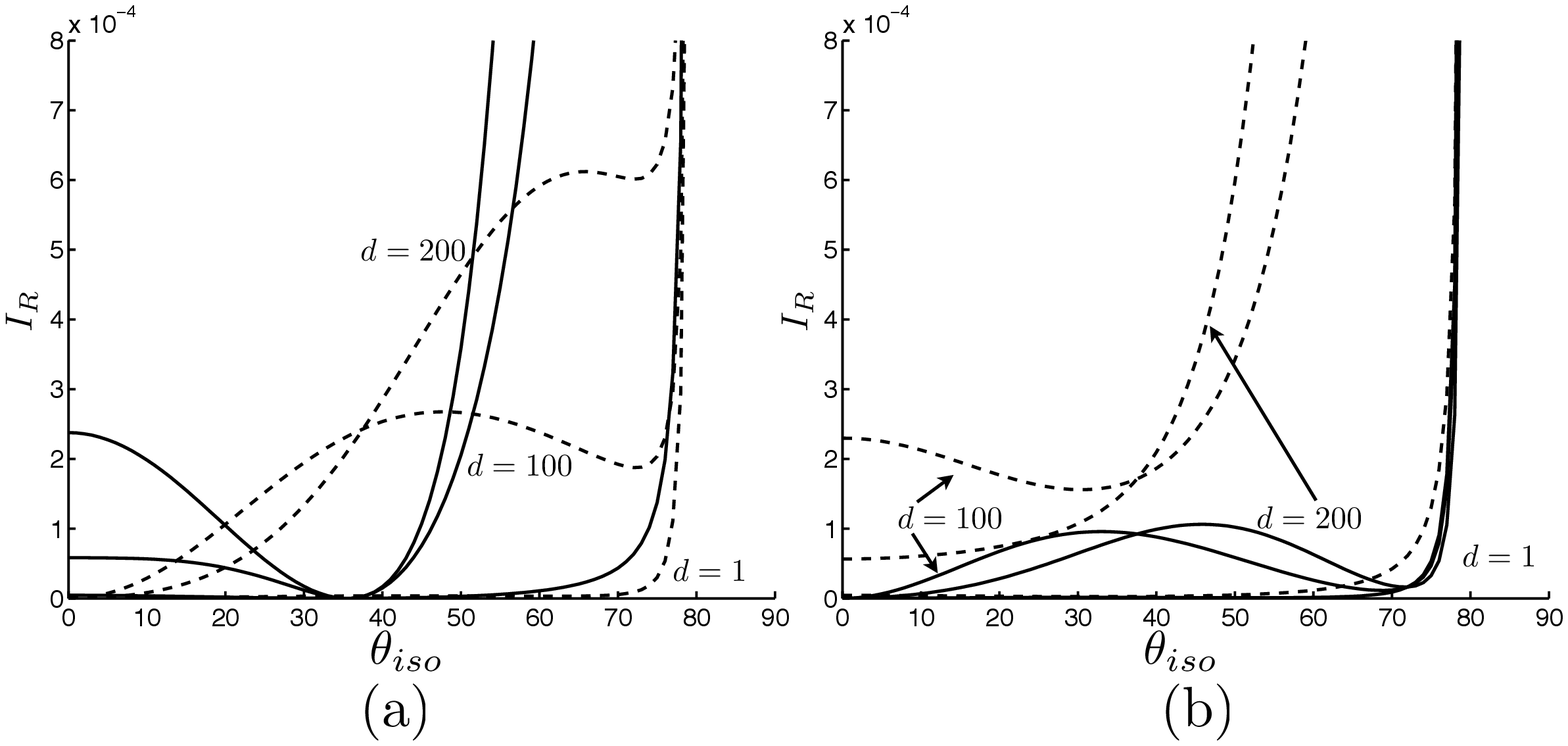}
\caption{Theory for reflected intensity ratio ($I_R$) from an ice-water-ice sandwich grown in an experimental ice growth cell; $[\theta_{\alpha}^1,\theta_{\beta}^1,\theta_{\gamma}^1]=[103\degree,131\degree,45\degree]$ and $[\theta_{\alpha}^2,\theta_{\beta}^2,\theta_{\gamma}^2]=[89.7\degree,88.5\degree,1.5\degree]$.  Intensity ratio is plotted for varying water layer thicknesses, $d=1,100,200$ nm.  In both cases solid curves correspond with reflected \emph{o} polarization and dashed curves with the \emph{e} polarization.  (a) The incident beam is \emph{o} polarized $[E^{o}_i,E^{e}_i]=[1,0]$.  (b) The incident beam is \emph{e} polarized $[E^{o}_i,E^{e}_i]=[0,1]$.  In both cases the mixing of polarization states is clear.}
\label{fig:samp}
\end{figure}

This is the extent to which we can easily pursue the problem algebraically.  Calculating an experimentally useful value, such as the reflected flux density (i.e. $I_r=\mathbf{E}_{r}^{tot} \cdot \mathbf{E}_{r}^{tot*}/2$), involves taking the complex conjugate of $\mathbf{E}_{r}^{tot}$ and solving for the reflected intensity by working through the full algebraic expressions.  This quickly becomes quite cumbersome as can be seen from the full expression in \eref{eq:Mn} for $M_n^{ref}$  but is easily facilitated by a symbolic or matrix based mathematical computer interface.  Examples of intensity ratio (i.e. $I_R=\mathbf{E}_{r}^{tot} \cdot \mathbf{E}_{r}^{tot*}/(\mathbf{E}_{i}\cdot \mathbf{E}_{i}^*$)) are plotted for reflection from a basal plane, as a function of incidence angle (\Fref{fig:90900}) and as a function of grain boundary film thickness (\Fref{fig:dbasal}).  For other orientations plots of intensity ratio illustrate some interesting characteristics.  For special crystallographic orientations (Figures~\ref{fig:90900},\ref{fig:90090}) no polarization mixing takes place.  However, when a particular polarization state within the crystal is parallel with the \emph{p} polarization in the isotropic layer, a Brewster-like angle exists for the system.  In \fref{fig:90900} this Brewster angle is present for the \emph{e} polarized wave, and conversely, in \fref{fig:90090} the \emph{o} polarized incident beam has an angle of zero reflection. For systems with less crystallographic symmetry (\Fref{fig:454590}) the mixing of polarization states is clearly important.  An examination of intensity ratio as a function of $\hat{c}$-axis orientation (\Fref{fig:rotcax}) illustrates the relative importance of polarization mixing and points where symmetries preclude coupling. Theoretical curves for experimentally measured crystallography (\Fref{fig:samp}) show substantial polarization mixing, yet for certain orientations and polarizations effective Brewster-like angles emerge. 

A nearly identical analysis can be done in order to calculate the theoretically transmitted fields and intensity ratios associated with an isotropic layer sandwiched between anisotropic media.  The infinite sum is simply rewritten in terms of the transmitted waves,
\begin{eqnarray}
\label{eq:ETt} \mathbf{E}_{Tt}&=&(T_2\sum_{n=0}^{\infty}(\bar{R}e^{-i\delta})^nT_1)\mathbf{E}_ie^{i\omega t}\\
\label{eq:ETt1}&=&[T_2(I-\bar{R}e^{-i\delta})^{-1}T_1]\mathbf{E}_ie^{i\omega t},
\end{eqnarray}
where we can label the bracketed portion of the expression the \emph{n}-transmission matrix,  
\begin{multline}\label{eq:Mntrs}
M_n^{tr}= \left( \begin{array}{c}
{[t_{op}t_{po}'+t_{os}t_{so}']e^{i2\delta}+[\eta_5t_{po}'+\eta_6t_{so}']e^{i\delta}}
\\ { [t_{op}t_{pe}'+t_{os}t_{se}']e^{i2\delta}+[\eta_5t_{pe}'+\eta_6t_{se}']e^{i\delta}}
 \end {array} \right . \\
\left. \begin{array} {c} 
{[t_{ep}t_{po}'+t_{es}t_{so}']e^{i2\delta}+[\eta_7t_{po}'+\eta_8t_{so}']e^{i\delta}}\\
{[t_{ep}t_{pe}'+t_{es}t_{se}']e^{i2\delta}+[\eta_7t_{pe}'+\eta_8t_{se}']e^{i\delta}}\end{array} \right)\frac{1}{(e^{i2\delta}-\eta_1e^{i\delta}+\eta_2\eta_3)} .
\end{multline}
Again the variable quantities witidhin the matrix elements are given in Appendix C.  Ensuing calculations follow in analogy to what we have previously shown for reflection.  Whereas for our experimental system this is not of interest, there may be experimental systems in which the transmitted wave would be the appropriate observable.

\section{Conclusions}

We have derived explicit expressions for the reflection and refraction amplitude coefficients for ordinary and extraordinary polarized electromagnetic waves incident upon an interface between a uniaxially anisotropic and an isotropic material.  The orientation of the optical axes may be arbitrary with respect to a laboratory frame of reference.  Furthermore, the formulae are valid for the full range of incidence angle.    

Combining these results with \possessivecite{Lekner1991} earlier work allows us to model a three layer system of arbitrarily oriented uniaxial crystals sandwiching an isotropic layer.  A Jones-like matrix formulation is used to treat the electromagnetic wave propagation and the reflection and transmission matrices for an isotropic layer sandwiched between anisotropic materials are explicitly determined.   Light scattering from the interfaces between two grains in polycrystalline ice is an area with wide ranging implications in astrophysical and geophysical settings, but also serves as an ideal, transparent analogue for many materials \cite{Dash2006}.  Comparable systems may be present in layered ceramics \cite{Luo2008}, biological structures \cite{Parsegian2006}, or experimental tests of the theory of dispersion forces in such layered geometries \cite{vanBenthem2006}.  Presently we are engaged in a long term experimental test of the theoretical framework described here \cite{Thomson2005a}.  It is hoped that, in consequence, those interested in the structure of grain boundaries in other systems can make use of both the framework laid out in this paper and the incipient experimental findings.

\appendix
\section*{Appendix A}
\setcounter{section}{1}

Here we summarize \possessivecite{Lekner1991} results for the reflection and transmission coefficients for \emph{s} and \emph{p} waves incident onto an isotropic/anisotropic interface.  We have made small changes where apparent typographical errors were made in the original manuscript.
\begin{eqnarray}
\label{eq:rss} r_{ss}=\frac{(q_{iso}^+-q_e^+)A_lE^{e^+}_y-(q_{iso}^+-q_o^+)B_lE^{o^+}_y}{D_l}, \\
\label{eq:rsp} r_{sp}=\frac{2(q_{iso}^+ \cos{\theta_{iso}} + K \sin{\theta_{iso}})(A_lE^{e^+}_x-B_lE^{o^+}_x)}{D_l} \\
\label{eq:tso} t_{so}=\frac{-2q_{iso}^+B_l}{D_l}\;\;\;\text{and} \\
\label{eq:tse} t_{se}=\frac{2q_{iso}^+A_l}{D_l},
\end{eqnarray}
where  $\theta_{iso}$ is the incident angle in the isotropic layer.  The subscripts $p,s,o$ and $e$ denote the particular amplitude coefficient, the $q$'s are the wave vector's $z$ components, and 
\begin{eqnarray}
\label{eq:lekA} A_l=(q_o^{+}+q_{iso}^++K \tan{\theta_{iso}})E_x^{o^+}-KE^{o^+}_z, \\
\label{eq:lekB} B_l=(q_e^{+}+q_{iso}^++K \tan{\theta_{iso}})E_x^{e^+}-KE^{e^+}_z, \\
\label{eq:lekD}D_l=(q_{iso}^++q_e^{+})A_lE^{e^+}_y-(q_{iso}^++q_o^{+})B_lE^{o^+}_y.
\end{eqnarray}
\noindent \Eref{eq:tse} has been altered after \citeasnoun{Lekner1992b} where he pointed out a misprint of the sign in \citeasnoun{Lekner1991}.   The coefficients for the incident \emph{p} wave are:
\begin{eqnarray}
\label{eq:rpp} r_{pp}= \frac{2(q_{iso}^+ \cos{\theta_{iso}} + K \sin{\theta_{iso}})F_l}{D_l \cos{\theta_{iso}}}-1, \\
\label{eq:rps} r_{ps}= \frac{2(q_{iso}^+ \cos{\theta_{iso}} + K \sin{\theta_{iso}})(q_e^{+}-q_o^{+})E^{o^+}_yE^{e^+}_y}{D_l} ,\\
\label{eq:tpo} t_{po}=\frac{2(q_{iso}^+ \cos{\theta_{iso}} + K \sin{\theta_{iso}})(q_{iso}^++q_e^{+})E^{e^+}_y}{D_l}\;\;\;\text{and} \\
\label{eq:tpe} t_{pe}=\frac{-2(q_{iso}^+ \cos{\theta_{iso}} + K \sin{\theta_{iso}})(q_{iso}^++q_o^{+})E^{o^+}_y}{D_l}, 
\end{eqnarray}
where $F_l=[(q_{iso}^++q_e^{+})E^{o^+}_xE^{e^+}_y-(q_{iso}^++q_o^{+})E^{o^+}_yE^{e^+}_x]$.  Here \eref{eq:rpp} differs from \citeasnoun{Lekner1991}, where we infer there was a typographical error.

\appendix
\section*{Appendix B}
\renewcommand{\theequation}{B-\arabic{equation}}
\setcounter{section}{1}

The full expressions for the amplitude coefficients associated with the ordinary wave at an interface with an isotropic material are,
\begin{multline}\label{eq:roo}
r_{oo}=\\
\frac{E_y^{o^+}(q_{iso}^+-q_o^{+})[E_x^{e^-}(A-q_{e}^{-})+KE_z^{e^-}]-E_y^{e^-}(q_{iso}^+-q_{e}^{-})[E_x^{o^+}(A-q_o^{+})+KE_z^{o^+}]}{B},
\end{multline}
where, $B\equiv E_y^{e^-}(q_{iso}^+-q_{e}^{-})[E_x^{o^-}(A-q_{o}^{-})+KE_z^{o^-}]-E_y^{o^-}(q_{iso}^+-q_{o}^{-})[E_x^{e^-}(A-q_{e}^{-})+KE_z^{e^-}]$ and $A\equiv q_{iso}^++K\tan{\theta_{iso}^+}$,
\begin{multline}\label{eq:roe}
r_{oe}=\\
\frac{E_y^{o^-}(q_{iso}^+-q_{o}^{-})[E_x^{o^+}(A-q_{o}^{+})+KE_z^{o^+}]-E_y^{o^+}(q_{iso}^+-q_{o}^{+})[E_x^{o^-}(A-q_{o}^{-})+KE_z^{o^-}]}{B}.
\end{multline}
\begin{multline}\label{eq:tos}
t_{os}=\\
B^{-1}[E_y^{o^+}\{E_y^{e^-}(q_{o}^{+}-q_{e}^{-})[E_x^{o^-}(A-q_{o}^{-})+KE_z^{o^-}]-E_y^{o^-}(q_{o}^{+}-q_{o}^{-})[E_x^{e^-}(A-q_{e}^{-})+KE_z^{e^-}]\}+\\
E_y^{e^-}E_y^{o^-}(q_{e}^{-}-q_{o}^{-})[E_x^{o^+}(A-q_{o}^{+})+KE_z^{o^+}]]
\end{multline}
\begin{multline}\label{eq:top}
t_{op}=\\
(B\cos{\theta_{iso}})^{-1}\{E_y^{o^-}(q_{iso}^+-q_{o}^{-})[-E_x^{e^-}E_x^{o^+}(q_{o}^{+}-q_{e}^{-})+K(E_x^{e^-}E_z^{o^+}-E_x^{o^+}E_z^{e^-})]+\\
E_y^{e^-}(q_{iso}^+-q_{e}^{-})[E_x^{o^-}E_x^{o^+}(q_{o}^{+}-q_{o}^{-})+K(E_z^{o^-}E_x^{o^+}-E_x^{o^-}E_z^{o^+})]+\\
E_y^{o^+}(q_{iso}^+-q_{o}^{+})[E_x^{e^-}E_x^{o^-}(q_{o}^{-}-q_{e}^{-})+K(E_x^{o^-}E_z^{e^-}-E_x^{e^-}E_z^{o^-})]\}
\end{multline}

For an incident wave polarized in the extraordinary direction the coefficients have a similar form; 
\begin{multline}\label{eq:ree}
r_{ee}=\\
\frac{E_y^{o^-}(q_{iso}^+-q_{o}^-)[E_x^{e^+}(A-q_{e}^+)+KE_z^{e^+}]-E_y^{e^+}(q_{iso}^+-q_{e}^{+})[E_x^{o^-}(A-q_{o}^-)+KE_z^{o^-}]}{B}
\end{multline}
\begin{multline}\label{eq:reo}
r_{eo}=\\
\frac{E_y^{e^+}(q_{iso}^+-q_{e}^{+})[E_x^{e^-}(A-q_{e}^-)+KE_z^{e^-}]-E_y^{e^-}(q_{iso}^+-q_{e}^-)[E_x^{e^+}(A-q_{e}^{+})+KE_z^{e^+}]}{B}
\end{multline}
\begin{multline}\label{eq:tes}
t_{es}=\\
B^{-1}[E_y^{e^+}\{E_y^{e^-}(q_{e}^{+}-q_{e}^{-})[E_x^{o^-}(A-q_{o}^{-})+KE_z^{o^-}]-E_y^{o^-}(q_{e}^+-q_{o}^-)[E_x^{e^-}(A-q_{e}^{-})+KE_z^{e^-}]\}+\\
E_y^{e^-}E_y^{o^-}(q_{e}^{-}-q_{o}^{-})[E_x^{e^+}(A-q_{e}^{+})+KE_z^{e^+}]]
\end{multline}
\begin{multline}\label{eq:tep}
t_{ep}=\\
(B\cos{\theta_{iso}})^{-1}\{E_y^{e^-}(q_{iso}^+-q_{e}^{-})[E_x^{e^+}E_x^{o^-}(q_{e}^{+}-q_{o}^{-})+K(E_x^{e^+}E_z^{o^-}-E_x^{o^-}E_z^{e^+})]+\\
E_y^{o^-}(q_{iso}^+-q_{o}^{-})[-E_x^{e^-}E_x^{e^+}(q_{e}^{+}-q_{e}^{-})+K(E_z^{e^+}E_x^{e^-}-E_x^{e^+}E_z^{e^-})]+\\
E_y^{e^+}(q_{iso}^+-q_{e}^{+})[E_x^{e^-}E_x^{o^-}(q_{o}^{-}-q_{e}^{-})+K(E_x^{o^-}E_z^{e^-}-E_x^{e^-}E_z^{o^-})]\}
\end{multline}

\appendix
\section*{Appendix C}
\renewcommand{\theequation}{C-\arabic{equation}}
\setcounter{section}{1}

Although any computation utilizing the presented theory is most efficiently done utilizing linear algebra, the variable quantities that compose the elements of the $n$-reflection matrices ($M_n^{ref}$ and $M_n^{tr}$) are presented here for completeness.  
\begin{eqnarray}
\label{eq:eta} \eta_1 &\equiv& \bar{R}_{11}+\bar{R}_{22}\\
\label{eq:eta2}\eta_2&\equiv &r_{ps}'r_{sp}'-r_{pp}'r_{ss}'\\
\label{eq:eta3} \eta_3 &\equiv& r_{ps}r_{sp}-r_{pp}r_{ss}\\
\label{eq:eta4}\eta_4&\equiv&r_{pp}t_{so}-r_{sp}t_{po}\\
\label{eq:eta5}\eta_5&\equiv&\bar{R}_{12}t_{os}-\bar{R}_{22}t_{op}\\
\label{eq:eta6}\eta_6&\equiv&\bar{R}_{21}t_{op}-\bar{R}_{11}t_{os}\\
\label{eq:eta7}\eta_7&\equiv&\bar{R}_{12}t_{es}-\bar{R}_{22}t_{ep}\\
\label{eq:eta8}\eta_8&\equiv&\bar{R}_{21}t_{ep}-\bar{R}_{11}t_{es}\\
\label{eq:zeta1} \zeta_1 &\equiv &r_{pp}'t_{ep}+r_{sp}'t_{es}\\
\label{eq:zeta2}\zeta_2&\equiv&r_{ps}'t_{ep}+r_{ss}'t_{es}\\
\label{eq:zeta3} \zeta_3 &\equiv &r_{ss}t_{po}-r_{ps}t_{so}\\
\label{eq:zeta4}\zeta_4&\equiv&r_{pp}'t_{op}+r_{sp}'t_{os}\\
\label{eq:zeta5} \zeta_5 &\equiv &r_{ps}'t_{op}+r_{ss}'t_{os}\\
\label{eq:zeta6}\zeta_6&\equiv&r_{ss}t_{pe}-r_{ps}t_{se}\\
\label{eq:zeta7} \zeta_7 &\equiv &r_{pp}t_{se}-r_{sp}t_{pe}.
\end{eqnarray}
Where the subscripted $\bar{R}$'s are the matrix elements of the previously defined $\bar{R}$,
\begin{equation}
\label{eq:Rbar}\bar{R}\equiv R_3R_2= \begin{pmatrix}r_{pp}'r_{pp}+r_{ps}'r_{sp}& r_{pp}r_{sp}'+r_{sp}r_{ss}' \\ r_{pp}'r_{ps}+r_{ps}'r_{ss} & r_{ps}r_{sp}'+r_{ss}'r_{ss} \end{pmatrix}.
\end{equation}

\section*{Acknowledgements}
We are grateful to M.L. Spannuth for reading an early draft of this manuscript. We thank the Leonard X. Bosack and Bette M. Kruger foundation, the US National Science Foundation (No. OPP0440841), the Department of Energy (No. DE-FG02-05ER15741), the Helmholtz Gemeinschaft Alliance, ``Planetary Evolution and LifeÓ, and Yale University for generous support of this research. Additionally, JSW acknowledges the support of NORDITA, the Royal Institute of Technology, Stockholm University and the Wenner-Gren Foundation all in Stockholm, Sweden.

\small\it{This is an author-created, un-copyedited version of an article accepted for publication in Journal of Physics: Condensed Matter. IOP Publishing Ltd is not responsible for any errors or omissions in this version of the manuscript or any version derived from it. The definitive publisher authenticated version is available online at \href{http://www.iop.org/EJ/abstract/0953-8984/21/19/195407}{http://www.iop.org/EJ/abstract/0953-8984/21/19/195407}}.
\section*{References}
\bibliographystyle{jphysicsB}
\bibliography{bib_layerscatter}
   
\end{document}